\def\spose#1{\hbox to 0pt{#1\hss}}
\def\simlt{\mathrel{\spose{\lower 3pt\hbox{$\mathchar"218$}}
     \raise 2.0pt\hbox{$\mathchar"13C$}}}
\def\simgt{\mathrel{\spose{\lower 3pt\hbox{$\mathchar"218$}}
     \raise 2.0pt\hbox{$\mathchar"13E$}}}
\def\simpropto{\mathrel{\spose{\lower 3pt\hbox{$\mathchar"218$}}
     \raise 2.0pt\hbox{$\propto$}}}

\documentclass[useAMS,usenatbib]{mn2e}
\usepackage{amsmath,graphicx,bm}
\title{An Improved Method for 21cm Foreground Removal}
\author[Adrian Liu, Max Tegmark, Judd Bowman, Jacqueline Hewitt, Matias Zaldarriaga]{Adrian Liu$^{1}$\thanks{E-mail:
acliu@mit.edu}, Max Tegmark$^{1}$, Judd Bowman$^{2,3}$,Jacqueline Hewitt$^{1}$, Matias Zaldarriaga$^{4,5}$\\
$^{1}$Dept. of Physics and MIT Kavli Institute, Massachusetts Institute of Technology, 77 Massachusetts Ave., Cambridge, MA 02139, USA\\
$^{2}$Dept. of Physics, California Institute of Technology, Pasadena, CA 91125, USA \\
$^{3}$Hubble Fellow \\
$^{4}$Harvard-Smithsonian Center for Astrophysics, 60 Garden St., Cambridge, MA 02138, USA\\
$^{5}$Jefferson Laboratory of Physics, Harvard University, Cambridge, MA 02138, USA}
\date{March 25, 2009}
\begin{document}

\pagerange{\pageref{firstpage}--\pageref{lastpage}} \pubyear{2008}

\maketitle

\begin{abstract}
$21\,\textrm{cm}$ tomography is expected to be difficult in part because of serious foreground contamination.  
Previous studies have found that line-of-sight approaches are capable of cleaning foregrounds to an acceptable level on large spatial scales, but not on small spatial scales.  
In this paper, we introduce a Fourier-space formalism for describing the line-of-sight methods, and use it to introduce an improved new method for $21\,\textrm{cm}$ foreground cleaning.  Heuristically, this method involves fitting foregrounds in Fourier space using weighted polynomial fits, with each pixel weighted according to its information content.  We show that the new method reproduces the old one on large angular scales, and gives marked improvements on small scales at essentially no extra computational cost.
\end{abstract}


\section{Introduction}
\label{intro}
Neutral hydrogen tomography is emerging as a promising new probe of the epoch of reionization and cosmology. By taking advantage of the $21\,\textrm{cm}$ hyperfine transition, neutral hydrogen tomography in principle allows one to map the distribution of hydrogen over a large range of redshifts, some of which are accessible through no other observational probes.  For example, neutral hydrogen tomography can potentially provide the only measure of the universe's expansion history, thermal history, as well as its clustering growth during the so-called dark ages.  Furthermore, the dramatic increase in the volume that can be mapped by the technique could enable precision tests of inflation, including stronger constraints on the spectral index of inflationary seed fluctuations, the running of the index, and small-scale non-Gaussianity \citep{Matt3, Santos2, Wyithe, juddjackiemiguel1, Yi}.  Neutral hydrogen tomography has also been predicted to be a sensitive probe of other parameters such as neutrino masses and the dark energy equation of state, either through power spectrum measurements \citep{Yi} or other probes such as $21\,\textrm{cm}$ lensing tomography \citep{Matiaszahn,Metcalf1,Metcalf2}.

\begin{figure}
\centering
\includegraphics[width=0.48\textwidth,trim=3.5cm 8.5cm 3.75cm 8cm, clip]{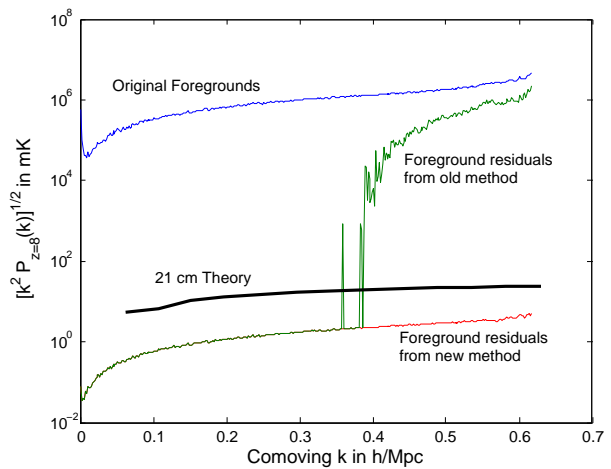}
\caption{2D power spectra of foregrounds and foreground residuals using the ``old method" \citep{Judd08,us} and the ``new method" (this paper).  At low-$k$ the two methods give identical results, while at high-$k$ the new method does much better.  Sudden spikes in the foreground residuals occur only with the old method.}
\label{master}
\end{figure}

Despite its promise, a number of challenges must be overcome before neutral hydrogen tomography becomes a reality.  One serious problem is the issue of foreground contamination.  A variety of astrophysical sources, including unresolved extragalactic points sources, resolved point sources, and galactic synchrotron radiation, will contribute contaminants with brightness temperature on the order of hundreds of Kelvins.  This will dominate the cosmological signal (which is expected to be on the order of mK), and so robust foreground subtraction techniques will be essential.

Previous studies have examined the feasibility of foreground subtraction in neutral hydrogen tomography, and have generally found that variations of the line-of-sight approach pioneered by \citet{ZFH}, \citet{xiaomin}, and \citet{Matt3} may be able to clean out foreground contamination to an acceptable degree, although instrumental effects such as noise may compromise the quality of the cleaned maps.  \citet{xiaomin,Judd08,lofar2,nusserforegrounds,Harker,us} performed simulations that included fiducial models of these effects, and found reasonably encouraging results.  It should also be noted that many of these instrumental effects, though serious, represent problems that are decoupled from the foreground subtraction challenge.  As discussed in \citet{us}, any \emph{linear} subtraction algorithm leaves the noise contribution to the power spectra unaffected, and so noise bias removal can be dealt with separately.  For instance, instrumental noise bias can be removed from power spectra by cross-correlating maps made from data taken at different times.  Whether the results are ultimately acceptable for Epoch of Reionization science will be difficult to answer until experimental data is obtained.

In any case, it is important to consider a wide variety of possible foreground subtraction algorithms, and in this paper we propose a new variation on the traditional line-of-sight methods.  Specifically, we describe a cleaning algorithm that (unlike most\footnote{\citet{ZFH} and \citet{nusserforegrounds} are exceptions and consider algorithms for Fourier space subtraction.} proposals) is implemented in Fourier space.  As we discuss in Section~\ref{new}, this allows one to completely sidestep any problems that may arise from the frequency dependence of an instrument's beam, which was previously the limiting factor in the quality of foreground cleaning at high-wavenumber spatial Fourier modes \citep{Judd08,us}.  The increase in performance at such wavenumbers can be easily seen in figure \ref{master}, where we have taken simulated data from a single frequency slice ($\nu = 158.73\,\textrm{MHz}$, corresponding to a 21-cm signal coming from $z=8$) and plotted $[k^2 P_{2D} (k)]^{1/2}$, where $P_{2D} (k)$ refers to the two-dimensional spatial power spectrum.  The quantity $[k^2 P_{2D} (k)]^{1/2}$ can be thought of as the fluctuation level as a function of scale, and is exactly analogous to $\delta_T / T \propto [\ell^2 C_{\ell}]^{1/2}$ in cosmic microwave background experiments.

The rest of the paper is organized as follows.  In Section~\ref{old} we review the old method used in \citet{Judd08,us}, and in~Section \ref{fourierdescription} we recast it as an algorithm in Fourier-space.  The Fourier-space description is then used to introduce our new method in Section~\ref{new}.  We conclude in Section \ref{conc}.
\section{Review of Old Method}
\label{old}
In general, the data collected from a typical 21-cm tomography experiment can be thought of as populating a ``data cube": stacks of 2D images separated by redshift or frequency.  Along the transverse directions, the axes are usually labeled in one of three ways:
\begin{enumerate}
\item Real-space coordinates $\theta_x$ and $\theta_y$.  In this case the data cube is a literal map of 21-cm emission and foreground contaminants.
\item Interferometer coordinates $u$ and $v$.  Under the correct convention, these are simply the Fourier-conjugate coordinates to $\theta_x$ and $\theta_y$.  The data cube is a stack of 2D maps in Fourier space.
\item Fourier-space coordinates $k_x$ and $k_y$.  These are the Fourier-conjugate coordinates to the physical lengths $x$ and $y$.  Up to factors of $2\pi$ (depending on one's Fourier convention), $(k_x, k_y) \sim (u,v) / D_M$, where $D_M$ is the transverse comoving distance.
\end{enumerate}
In a typical experiment the data (in the form of visibilities) are collected in $uv$-coordinates, while the results are presented in either real-space coordinates (in the case of sky maps) or in Fourier-space coordinates (in the case of power spectra).  Foreground removal is often done in either real-space coordinates (as demonstrated in \citet{xiaomin,Judd08,lofar2,us}) or in $uv$-space (as done in \citet{ZFH,nusserforegrounds}, and as we propose in this paper).

We first review the real-space removal algorithms.  The fundamental idea behind all such algorithms is 
the fact that the 21-cm signal is expected to oscillate rapidly with frequency while the relevant foreground contaminants are spectrally smooth.  The contaminants along a given line-of-sight can therefore be separated from the signal by plotting the flux as a frequency and subtracting off a smooth component (such as a low-order polynomial) from the total signal.  What remains is the cosmological signal and a (hopefully small) residual contamination.

Previous studies have simulated the aforementioned real-space algorithms and have estimated the level of residual contamination that can be expected for current experiments \citep{lofar2,Judd08} as well as how the residuals depend on the properties of a generic interferometer \citep{us}.  Although these papers have highlighted the fact that the quality of foreground subtraction is highly dependent on a large number of parameters (both instrumental and those pertaining to data analysis), they also suggest that the qualitative behavior is rather generic.  In what follows we examine the qualitative behavior that emerges, emphasizing the various features and their mathematical origin.

\begin{figure}
\centering
\includegraphics[width=0.5\textwidth,trim=1.5cm 6cm 1.5cm 7cm, clip]{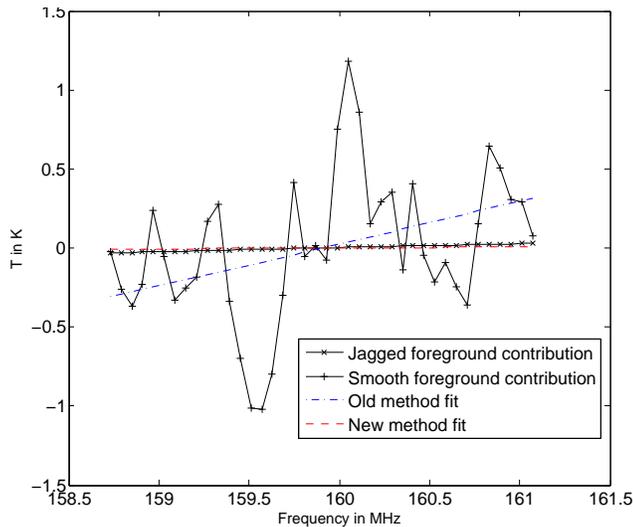}
\caption{The spectrum of a typical real-space pixel as seen by an interferometer.  The instrument introduces a jerky dependence on frequency even though the foregrounds are intrinsically smooth.  The total signal is the sum of a smooth component (black curve with ``x" markers) coming from the central parts of the $uv$-plane and a jagged component (black curve with ``+" markers) from the outer parts of the plane.  The blue/dot-dashed line gives the foreground fit using the old method, while the red/dashed line gives the analogous real space ``fit" using the new method (see Section~\ref{new} for details).  The means of each curve have been artificially removed for clarity.}
\label{realspace}
\end{figure}

Consider the spectra shown in figure \ref{realspace}.  The sum of the two black curves show the frequency dependence of a single pixel in real-space coordinates (i.e. the frequency dependence of a particular line-of-sight), as seen by a typical $21\,\textrm{cm}$ tomography interferometer\footnote{We use the Murchison Widefield Array as our fiducial model for the simulations in this paper (see \citet{us} for details), but it should be noted that the algorithm we propose in Section~\ref{new} can be applied to data collected by any interferometric configurations.}.  This total spectrum (formed from the sum of the two black curves) contains \emph{foregrounds only}, with no noise\footnote{In our simulations, we neglect instrumental noise.  This represents no loss of generality because our subtraction algorithms are linear (please see Section \ref{intro} or \citet{us} for details).} or cosmological signal.  Since the foregrounds are known (and are simulated\footnote{The simulation methodology used in this paper was the same as that used in \citet{us}, where point sources were independently generated in each pixel from source count distributions given in \citet{dimatteo1}.  Please see \citet{us} for details.}) to be spectrally smooth, this suggests that the rapid oscillations seen in the figure are \emph{caused by the instrument}.  This is bad news for the subtraction algorithm, as it means that simply fitting out the smooth component of a spectrum will leave residuals that can be confused with the cosmological signal.  Indeed, it can be seen from the figure that the fit seems rather poor.

\begin{figure}
\centering
\includegraphics[width=0.5\textwidth,trim=0cm 0cm 0cm 0cm, clip]{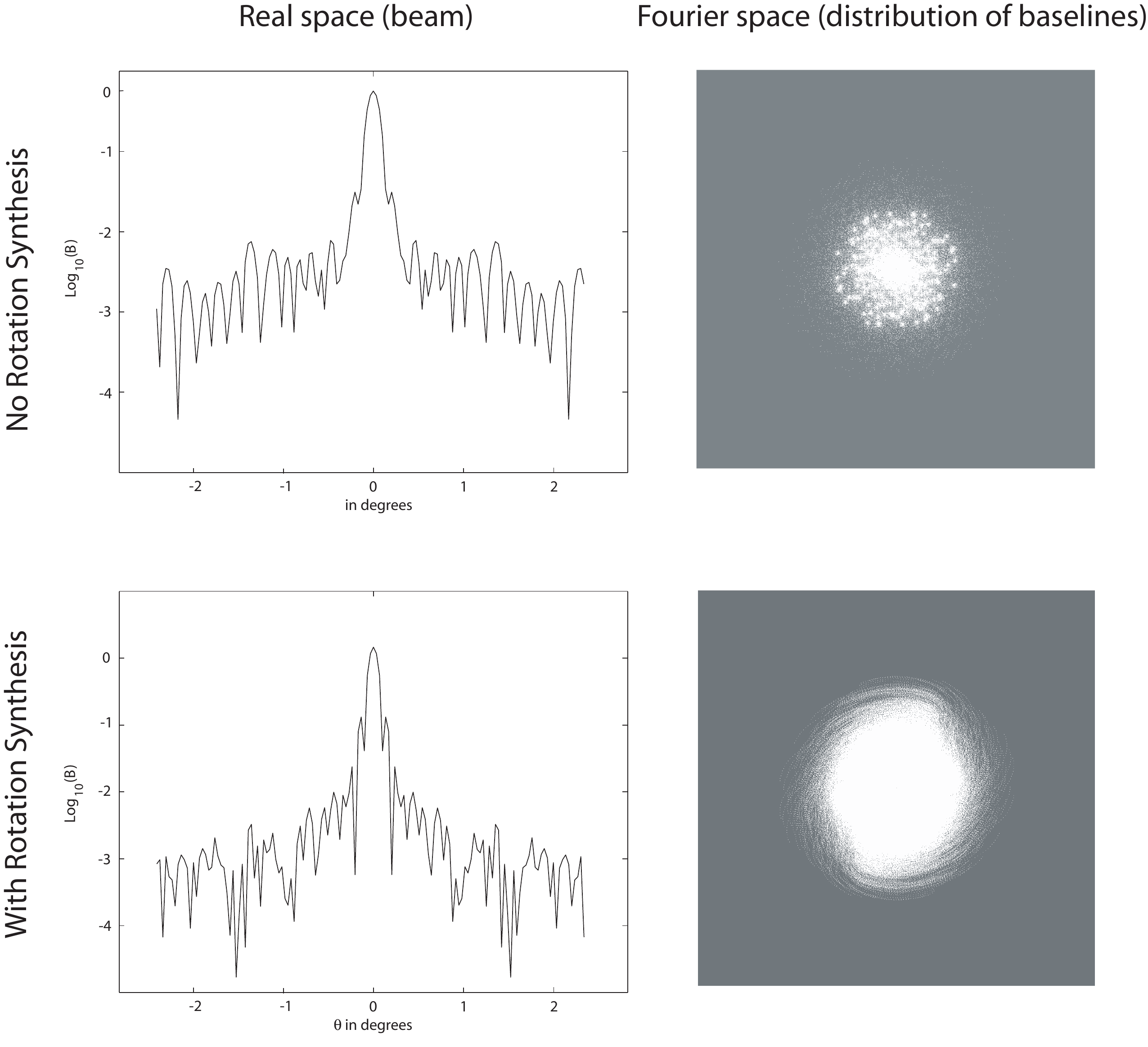}
   \caption{The left hand column shows sample beam profiles (a real-space description of the beam) while the right hand column shows the corresponding $uv$-distribution of baselines (a Fourier-space description of the beam). The top row illustrates an array with no rotation synthesis, while the bottom row shows an array with $6$ hours of rotation synthesis.  The real-space beams are normalized so that their peaks are at $1$.}
\label{pedfourier}
\end{figure}

One way of understanding the rapid oscillations is to consider the interferometer's beam in real-space.  The left-hand panel of figure \ref{pedfourier}, shows that the beam of a typical interferometer contains ``frizz" outside the central peak that oscillates rapidly with angle.  Since beam widths scale as $\lambda / D$, this angular oscillation translates into an oscillation in frequency, which is what is seen in figure \ref{realspace}.  Alternatively, the behavior of figure \ref{realspace} can be understood by considering the effect of an interferometer's beam in $uv$-space.  An interferometer samples pixels in the $uv$-plane, and with enough of these $uv$-pixels one can produce a real space image by Fourier transforming.  Thus, the spectrum of a single pixel in real space can be thought of as a linear combination of the spectra of different $uv$ pixels sampled by the interferometer.  Exactly which pixels are sampled depends on the layout of the interferometer in question, but in a typical 21-cm tomography experiment the $uv$ coverage is complete near the origin and drops off as one moves farther out.

In general, the foreground spectrum seen by an instrument can be considered the sum of two components: a component that is formed from a linear combination of $uv$-pixels where the interferometer's coverage is complete (i.e. the inner parts of the $uv$-plane), and a component that is formed from $uv$-pixels residing in parts of the $uv$-plane where coverage is sparse (i.e. the outer regions).  These components are shown using solid black lines in figure \ref{realspace}.  The line with ``x" markers (showing the part of the signal originating from the inner parts of the plane) is seen to be smooth, whereas the line with ``$+$" markers (showing the contribution from the outer parts) is what contributes the rapid oscillations.  (Note that this curve appears to have zero temperature only because we have artificially removed the mean of each curve for graphical clarity).  This decomposition explains why real-space pixel-by-pixel foreground subtraction algorithms have been shown to be adequate even though the fits themselves seem terrible at first sight.  Even though the smooth fits cannot subtract off the jerky component of the spectrum, they are capable of fitting out the smooth component that comes from the central parts of the Fourier plane.  Indeed, this is exactly what is seen in figure \ref{master}, where the low-$k$ parts of the power spectrum are cleaned effectively whereas the high-$k$ parts remain contaminated.  It is simply the case that by examining pixels in real space, one is viewing a ``bad" linear combination of pixels that mixes together the well-fit, central located $uv$-pixels with the outer $uv$-pixels where sparse baseline coverage results in jerky spectra that are badly fit.

\subsection{Fourier space description of decontamination}
\label{fourierdescription}
In the previous section, we examined how foreground fits of real-space pixels could be understood by considering the flux in each pixel as being a linear combination of different $uv$-pixels.  We now show that one can go further and perform the fits themselves in $uv$-space and get exactly the same results.  With slight modifications, this will lead to the discussion in Section~\ref{new} of an improved method for subtracting foregrounds at high $k$.

Consider the steps that must be taken to perform the foreground subtraction outlined above.  The data is collected by the interferometer in Fourier space i.e. in a $(u,v,\nu)$ data cube.  This data must then be Fourier transformed in the two transverse directions, giving an $x$-$y$-$\nu$ data cube.  Fitting is subsequently performed in the frequency direction.  Mathematically, we can express this as follows.  Let $\tilde{y}_{ij\alpha}$ represent the initial data cube, with the first two (Latin) indices being the two spatial indices and the last (Greek) index being the frequency index.  With no loss of generality, we can fold the first two indices into one and write $\tilde{y}_{j\alpha}$ instead.  In this notation, the Fourier transform can be written as
\begin{equation}
y_{k\alpha}=\sum_i \mathbf F_{ki} \tilde{y}_{i\alpha},
\end{equation}
where $\mathbf F$ is the Fourier matrix and $y$ is the real-space analog of $\tilde{y}$.  The fit in the frequency direction can be represented by yet another linear operator\footnote{Explicitly, for the case where one fits a polynomial of degree $m$, one has 
$\mathbf {G=X [X^t N^{-1} X]^{-1} X^t N^{-1}}$, where $\mathbf N$ is the noise covariance matrix
and $\mathbf X$ is an $n \times (m+1)$ matrix such that $\mathbf X_{ij}$ equals the frequency of the $i$th frequency channel taken to the $j$th power \citep{xiaomin}.} 
$\mathbf G$, and so we have
\begin{equation}
\label{sums}
\overline{y}_{k\beta}=\sum_\alpha \mathbf G_{\beta \alpha} y_{k \alpha} = \sum_{i,\alpha} \mathbf {G_{\beta \alpha} F}_{ki} \tilde{y}_{i\alpha}
\end{equation}
where $\overline{y}$ represents the fit.  In the last expression, note that $\mathbf G$ possesses only Greek indices whereas $\mathbf F$ only has Latin indices.  This means that the two operations performed in our algorithm -- the 2D spatial Fourier transform ($\mathbf F$) and the fitting in the frequency direction ($\mathbf G$) -- in fact commute, i.e., $\mathbf {FG=GF}$.

\begin{figure}
\centering
\includegraphics[width=0.45\textwidth,trim=0cm 0cm 0cm 0cm, clip]{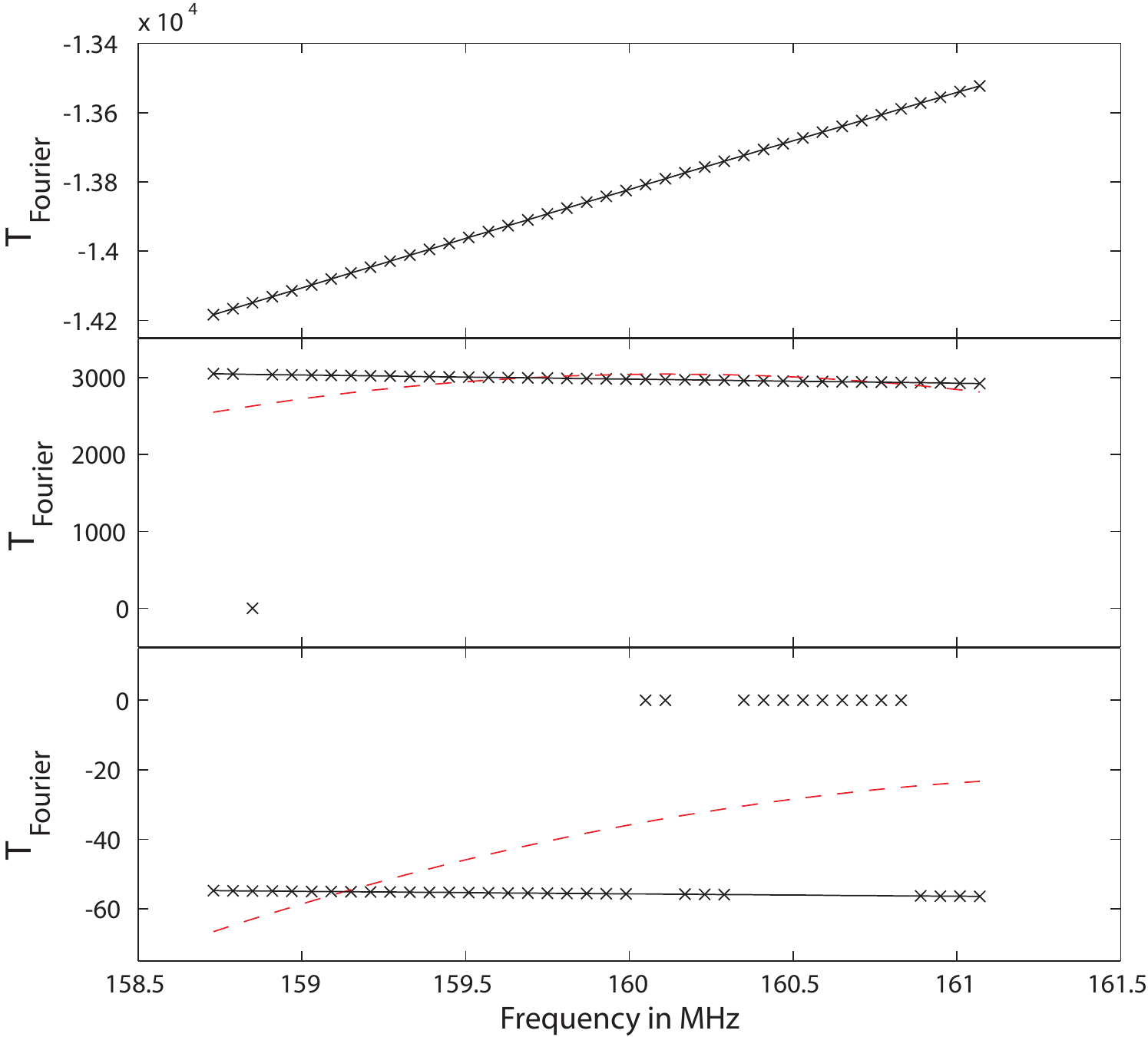}
\caption{Spectra of various $uv$ pixels from different parts of the plane.  From the top panel to the bottom panel, one is moving away from the origin.  It is clear that the data can be easily fit by low-order polynomials in the top panel, but that the old method of fitting (dashed red curves) becomes inadequate when baseline coverage begins to drop out.  The solid black curves show the fits done using the new method describe in Section~\ref{new}.}
\label{midfour}
\end{figure}

The fact that the Fourier transform commutes with the fitting means that we can perform the two operations in either order.  In other words, we can think of the foreground fitting and subtraction as taking place in Fourier space without changing any of the results (which is something that we have also verified numerically).  Viewing the process as a pixel-by-pixel fitting in $uv$-space reveals exactly why there exists such a vast difference between the quality of the cleaning at low-$k$ and at high-$k$, and why the transition between the two regimes appears as such an abrupt jump in the power spectra.  In figure \ref{midfour} we show typical spectra from different parts of the $uv$-plane. The top panel shows a typical pixel from the inner part of the plane.  The spectrum is plotted using so-called uniform weighting, so that in every Fourier pixel the interferometer acts as an on/off switch: the interferometer imposes a weighting of $0$ to a pixel if no baselines fall in that pixel, and a weighting of $1$ otherwise (regardless of how many baselines are binned into that pixel).  It is evident that a simple polynomial fit does extremely well.

On the other hand, when one moves out to regions of the $uv$-plane where baseline coverage becomes sparse, the fit becomes poor.  A glance at the bottom two panels of figure \ref{midfour} makes the problem clear -- when coverage is sparse, at certain frequencies there is no baseline coverage, and a simple polynomial fit is unable to deal with this.  We emphasize that the trouble is not with incomplete Fourier coverage per se.  It is the fact that the incomplete coverage is changing with frequency.  In other words, foreground subtraction becomes poor in this regime because the frequency dependence of the beam (or ``mode-mixing", as emphasized in \citet{Judd08,us}) becomes important on these small (high-$k$) scales.  Note that even though this problem exists when the spectra are being fit in real space, it is not \emph{apparent} unless one fits in $uv$-space, where the pixels are ``good" linear combinations of the data.

\section{New method}
\label{new}
We now propose a slight modification to the foreground subtraction algorithm that evades the aforementioned problem.  From figure \ref{midfour}, one can see that an alternate way of phrasing the problem is to say that the old fitting algorithm, being mathematically equivalent to a fitting in real space, is unable to distinguish between pixels with no data and pixels with values that happen to be zero.  In $uv$-space, however, one can easily identify pixels with no baselines, and so one can simply skip frequencies where data is unavailable.  In fact, one can find the optimal fit (in the sense of having minimal r.m.s. errors) by employing an inverse-variance weighted fit.  In this scheme, the weight of each point in the least-squares sum is proportional to $N$, the number of baselines that are binned into a particular $uv$-pixel at a particular frequency.  This way, points with lower signal-to-noise are given less weight, and points with no data at all are given zero weight \footnote{It is important to emphasize that in this section, we use the term ``weight" to refer to the statistical weight that we give to a data point in the fit.  We are \emph{not} pre-multiplying the data with a weighting function.  In other words, while our fits assign different statistical weights to each data point, the data points themselves are not tampered with ahead of time and are simply ``uniformly pre-weighted" as described in Section~\ref{fourierdescription}.}

\begin{figure*}
\centering
\includegraphics[width=1.0\textwidth,trim=0cm 0cm 0cm 0cm, clip]{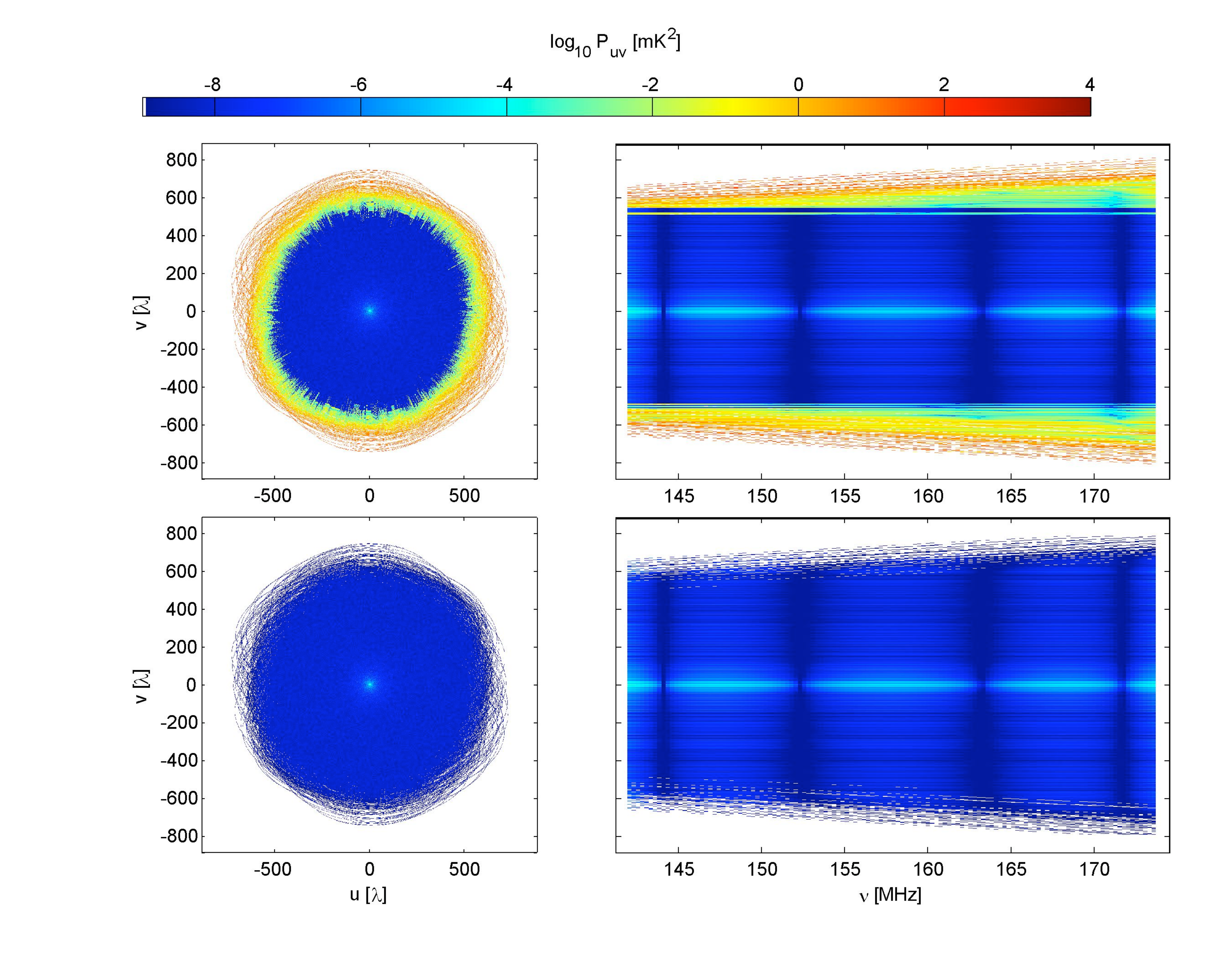}
\caption{Post-subtraction residuals shown in the $uv$-plane at $\nu=157\,\textrm{MHz}$ (left column) and as a function of frequency, taking a cut through the center of each $uv$-plane (right column).  This is done for the old method (top row) as well as for the new method (bottom row).  The new method does not offer any increase in performance at low-$k$, but avoids the large increase in residuals at high-$k$.}
\label{Judd}
\end{figure*}

In figure \ref{midfour} it can be seen that since missing frequencies are now given zero weight in the fit, one obtains excellent fits even for $uv$-pixels where baseline coverage is sparse.  This improves the subtraction of foregrounds at frequencies where there \emph{is} data, whereas at the skipped frequencies nothing has been compromised since no foregrounds were detected by the instrument in the first place.

The effect that the frequency-skipping has on the 2D power spectrum is shown in figure \ref{master}.  To be conservative, we have also tested our new algorithm using a completely independent pipeline with a different foreground model (for details, please see \citet{Judd08}).  The results from the second pipeline are shown in figure \ref{Judd}, and the fact that the results agree demonstrate the fact that $uv$-plane cleaning is generally applicable and not dependent on the foreground model.  Qualitatively, one can see that at low-$k$ there is no improvement from the old method because in that regime one is limited by the fact that simple low-order polynomials will not in general be perfect fits to the foregrounds, even though the foregrounds are smooth functions.  At high-$k$, however, one avoids the dramatic increase in post-subtraction foreground residuals, because previously the limitation at high-$k$ was the mode-mixing problem.  With our new method, the limiting factor is the ability of the fitting function to match the form of the foregrounds.  For example, the fact that the foreground residuals in figure \ref{master} are a constant factor ($\sim 10^6$) off from the original foregrounds regardless of scale (or equivalently, regardless of location on the $uv$-plane) means that the residuals are due entirely to the quality of the fit.  In other words, the residuals of one part in $\sim 10^6$ come from the fact that the second-order polynomials used in the fits to produce figure \ref{master} are good fits to the foregrounds only to one part in $\sim 10^6$.  With the chief limitation now being the fitting itself, one can in principle subtract foregrounds up to Fourier modes that correspond to the longest baselines, although as one is forced to skip many frequencies at high $k$, the signal-to-noise of the data is reduced.

As a weighting scheme that weights data points according to their information content, inverse-variance weighting not only gives higher signal-to-noise data points greater weight, but also automatically incorporates frequency-skipping, since the skipped frequencies are simply those with $N=0$ and therefore no information.  While both effects contribute to better foreground subtraction, we find frequency-skipping to be the dominant cause of this increase in performance.

It is important to note that whereas without the skipping of empty frequencies the transverse Fourier transform commuted with the fitting of the foregrounds, under the new scheme proposed here the two operations no longer commute.  This is because the frequencies of the pixels that need to be skipped require knowing the baseline distribution (which lives in $uv$-space) and therefore depends on the location of the $uv$-pixel being cleaned.  Mathematically, this means that in equation \ref{sums}, the fitting operator $\mathbf G$ acquires an extra $i$ (spatial) index and the two sums no longer commute.  The significance of this is that the fit can no longer be done in real-space.  To apply this new algorithm for foreground subtraction, one \emph{must} work in Fourier space.

However, while the skipping of frequencies in our fit dictates that we must \emph{work} in Fourier space, the improvements brought about by the new algorithm can still be \emph{seen} in real space.  Consider the dashed (red) fit in figure \ref{realspace}.  This fit was obtained by taking the $uv$-space fits generated by the new algorithm and Fourier transforming real space to give a real space ``fit".  It is clear from the figure that the new method does a much better job of tracking the behavior of the smooth foreground component.  On the other hand, the slope of the fit from the old method is biased by the jagged foreground contribution (which, remember, is an instrumental artifact that arises from incomplete baseline coverage), and does a worse job tracking the smooth foregrounds.

The fact that our new method traces the smooth foreground component better means that it can be used to get better estimates of the foregrounds themselves.  One simply Fourier transforms the fits produced by the new algorithm to get real-space, multi-frequency maps of the foregrounds.  Such maps will be of a higher quality than those that are simply imaged by the instruments.  This is because our new fitting algorithm can be interpreted as one where the missing frequencies are not so much skipped as interpolated over.  By fitting low-order polynomials over the frequencies where data is available, one is essentially deriving a foreground model that can be extrapolated to other frequencies.  Without missing frequencies in the spectra, the real space foreground maps will not have artificially jagged foreground components, and will therefore be a more accurate representation of the true foregrounds.

\section{Conclusions}
\label{conc}
In this paper, we have shown that there is an easy explanation for the increased foreground residuals at high-$k$: a frequency-dependent incompleteness of baseline coverage in the outer parts of the $uv$-plane makes the foregrounds in certain $uv$-pixels difficult to fit out using a simple unweighted polynomial fit.  The solution to this problem is to weight the fit so that frequencies with no information are given zero weight, while other frequencies are given an inverse-variance weighting.  As seen in figure \ref{master}, this allows foreground cleaning to be performed at much higher $k$, paving the way for higher quality power spectrum measurements in neutral hydrogen tomography.

\section*{Acknowledgments}
We  wish to thank Mike Matejek, Miguel Morales, Leslie Rogers, and Christopher Williams for helpful comments.  MT was supported by NSF grants AST-0134999 and AST-05-06556 as well as fellowships from the David and Lucile Packard Foundation and the Research Corporation.  MZ was supported by NASA NNG05GJ40G and NSF AST-05-06556 as well as the David and Lucile Packard, Alfred P. Sloan and John D. and Catherine T. MacArthur foundations.

\bibliographystyle{mn2e}
\bibliography{21cmps2}

\end{document}